\documentclass[a4paper, 10pt, conference]{ieeeconf}      

\IEEEoverridecommandlockouts                              

\usepackage{graphics} 
\usepackage{epsfig} 
\usepackage{booktabs}
\usepackage{mathtools}
\usepackage{tikz}

\newcommand\copyrighttext{%
  \footnotesize \textcopyright 2023 IEEE. Personal use of this material is permitted. Permission from IEEE must be obtained for all other uses, in any current or future media, including reprinting/republishing this material for advertising or promotional purposes, creating new collective works, for resale or redistribution to servers or lists, or reuse of any copyrighted component of this work in other works.}
\newcommand\copyrightnotice{%
\begin{tikzpicture}[remember picture,overlay]
\node[anchor=south,yshift=10pt] at (current page.south) {\fbox{\parbox{\dimexpr\textwidth-\fboxsep-\fboxrule\relax}{\copyrighttext}}};
\end{tikzpicture}%
}

\title{\LARGE \bf
Context-Aware Automated Passenger Counting Data Denoising
}

\author{Noëlie Cherrier$^{*}$, Baptiste Rérolle, Martin Graive, Amir Dib and Eglantine Schmitt$^{1}$
\thanks{$^*$Corresponding author}%
\thanks{$^{1}$Authors are with Citio, 54 Quai de la Rapée, 75012 Paris France.
        {\tt\small \{firstname\}.\{lastname\}@cit.io}}%
}

\begin{document}

\maketitle
\copyrightnotice
\thispagestyle{empty}
\pagestyle{empty}

\begin{abstract}

A reliable and accurate knowledge of the ridership in public transportation networks is crucial for public transport operators and public authorities to be aware of their network's use and optimize transport offering. Several techniques to estimate ridership exist nowadays, some of them in an automated manner. Among them, Automatic Passenger Counting (APC) systems detect passengers entering and leaving the vehicle at each station of its course. However, data resulting from these systems are often noisy or even biased, resulting in under or overestimation of onboard occupancy. In this work, we propose a denoising algorithm for APC data to improve their robustness and ease their analyzes. The proposed approach consists in a constrained integer linear optimization, taking advantage of ticketing data and historical ridership data to further constrain and guide the optimization. The performances are assessed and compared to other denoising methods on several public transportation networks in France, to manual counts available on one of these networks, and on simulated data.

\end{abstract}

\section{INTRODUCTION}

In a context of public transport networks development, knowledge of network ridership is a key issue for both transport operators and public authorities. Onboard occupancy is an extremely valuable indicator of ridership for adapting the transportation offer and ensuring a sufficient level of comfort for passengers. 

There are various approaches to compute onboard occupancy using different data sources, such as manual counts and ticketing data, the two most frequently used data sources. However, these two sources have flaws: manual counts only picture networks on one day of the week, typically a workday, and are expensive as they require human resources to cover the desired perimeter of the network. The use of ticketing data, that usually record only the entrance station, requires an Origin/Destination (O/D) reconstruction process to get alighting stations and therefore onboard occupancy. This method is moreover not exhaustive as it does not include fraud.

\begin{figure}
  \centering
  \includegraphics[width=1\linewidth]{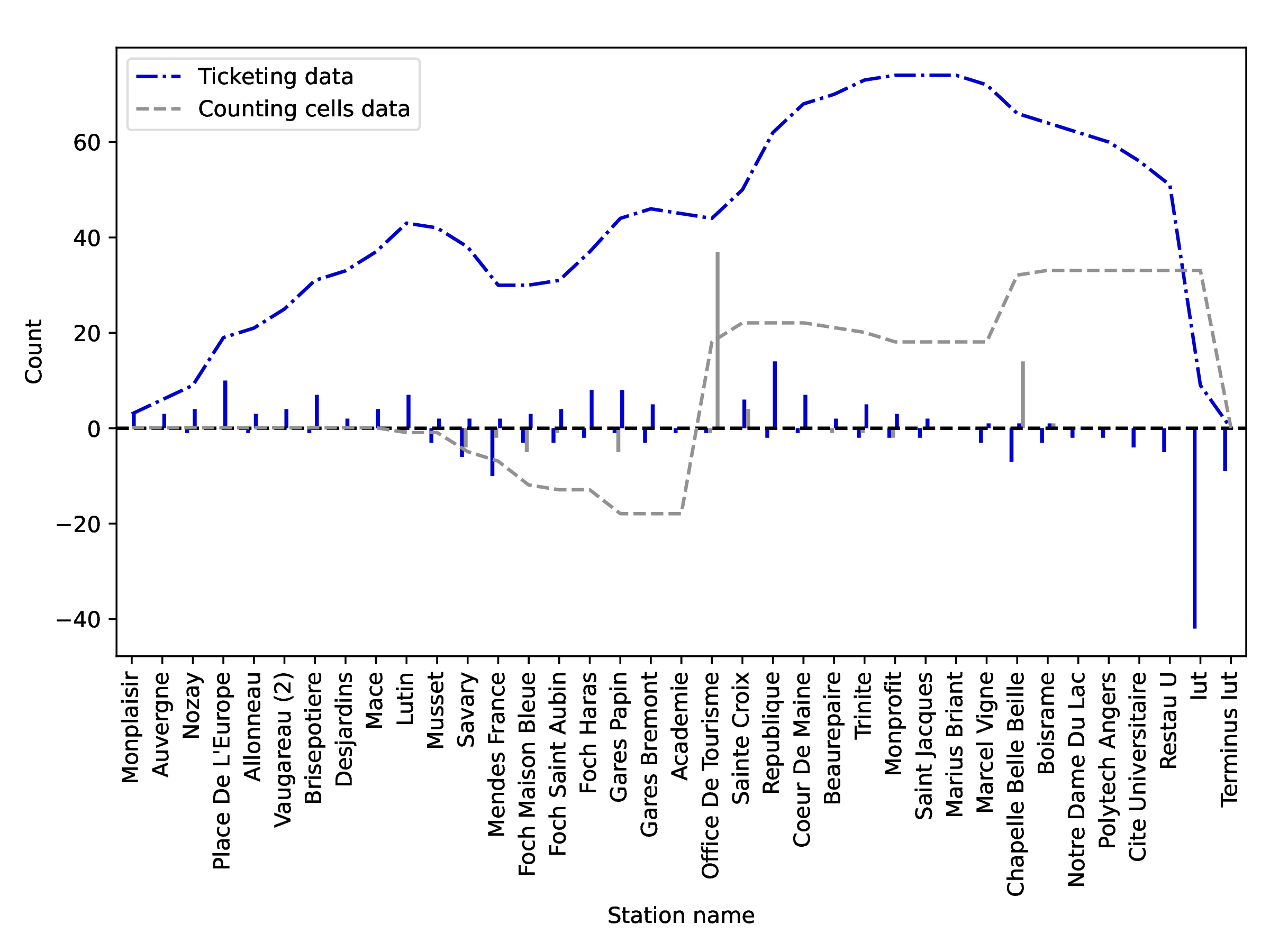}
  \caption{A course at Angers network. Due to biases in the counting cells measurements, occupancy computed from counting cells data is lower than the one computed from occupancy. It is even negative at the middle of the race which is not possible.}
  \label{fig:intro_course}
\end{figure}

The approach on which this paper focuses is the occupancy reconstruction using Automated Passenger Counting (APC) systems located onboard the vehicles (a review of which is found in \cite{kuchar2023passenger}). Some networks use computer vision to get the occupancy through image analysis, other networks use infrared sensors located above doors to count boardings and alightings at each stop \cite{grgurevic2022review}, etc. However, data from those sensors are noisy and have biases \cite{comparisonWLAN}. Fig.~\ref{fig:intro_course} illustrates several of these phenomena by tracking the occupancy along the successive stops of a course, on line 1 of the Irigo network (city of Angers). On this figure, the positive vertical bars represent boardings, the negative vertical bars represent alightings while the curves represent the occupancies. This course shows an occupancy computed from ticketing data that is higher than the one computed from APC data, while the latter is supposed to include fraud and therefore to be superior to the first one. Indeed, the boarding counts are overestimated and/or the alightings counts underestimated by the sensors. A second consequence of these sensor biases is observed in the middle of the course where counting cells occupancy is negative. This is an example of why APC data needs to be processed and cleaned to be reliable and exploitable at the course level.

This paper presents a context-aware method for denoising APC data based on a constrained integer linear optimization. Our contributions lie in the use of other available data such as ticketing data and vehicle capacity, in priors computations from historical data and in the performance evaluation of our approach. Indeed, results are tested on real data from several networks in France. They are compared to field studies and expert knowledge. Simulations are performed as well to ensure the robustness of the proposed approach. 

Section~\ref{sec:related_work} introduces the related work in this field, widening the subject to treatments made by APC systems providers. The proposed method is presented in Section~\ref{sec:proposed_method}, detailing the assumptions made and the optimization stages. Finally, comprehensive experiments on real-world data are conducted in Section~\ref{sec:experiments}.

\section{RELATED WORK}
\label{sec:related_work}

Raw sensor data is first processed online and in a technology-specific manner: computer vision techniques apply to video streams \cite{nasir2018automatic,hsu2020estimation,zhang2020tiny}, statistical methods to time-of-flight systems \cite{klauser2015tof}, among various other cleaning techniques \cite{amin2008automated,choi2017bi,nitti2020iabacus}. APC systems providers communicate their accuracy metrics based on the APC data resulting from these treatments. These are usually the only processings applied to get accurate APC data, although they do not ensure boarding and alighting counts consistency nor passenger flow conservation over an entire course. For instance, Patlins and Kunicina \cite{patlins2015new} take passenger flow conservation as granted and propose a photogrammetry approach for remote sensing of boarding and alighting passengers. Although recent technologies such as computer vision based systems are the most accurate, their drawbacks remain software support and most importantly data privacy \cite{grgurevic2022review}.

APC data is mostly used to reconstruct O/D matrices: the balancing or iterative proportional fitting methods permit to get an O/D matrix from a seed matrix (coming either from survey or ticketing data) and from APC counts that give the total number of boarding and alighting passengers \cite{ben1985alternative,lamond1981bregman,ji2015transit}. These methods make it possible to enforce passenger flow conservation but at the network level, since they are usually applied using aggregated data over some temporal range. In the proposed work, passenger flow conservation is guaranteed at the course level, so as to give a higher level of precision to operators.

A few articles propose methods to adjust a posteriori boarding and alighting counts on a network. Some of them acknowledge technical problems due to sensors' inaccuracy, and propose a deterministic process comprising correction factors followed by a rounding procedure \cite{furth2005making,barabino2014offline}. These works also enforce a threshold on negative occupancy below which a course should be rejected. 
These methods, based on expert knowledge, provide no guarantee on the distance between corrected counts and raw APC counts. 

Statistical methods have been proposed to sample valid passenger counts from prior observations, here raw APC data \cite{van1982consistent,amblard2023bayesian}. The principle is to assume counts follow a probability distribution constrained by passenger flow conservation. However, due to the statistical nature of these approaches, different results are obtained in each run, and a course with already valid counts may be modified in the process.

Besides, optimization-based approaches aim at minimizing the distance between observed and target counts. Several variants have been proposed, mainly differing by the chosen cost function. The most efficient choice for computational speed is $\ell_1$ minimization \cite{yin2017l1}. $\ell_2$ minimization has been used for vehicle conservation with traffic data \cite{vanajakshi2004loop}. More specifically, Kikuchi \textit{et al.} introduced fuzzy optimization to balance observed counts in a public transport: the principle is to authorize a range of possible values around the observed count \cite{kikuchi2006method}. On top on the cost function proposed in \cite{kikuchi2006method}, de Oña \textit{et al.} \cite{de2014adjustment} came up with a two-stage optimization to first reduce aberrant values, and then minimize the distances between observed and target counts according to a fuzzy cost function.

The proposed approach is built on top of the two-stage optimization introduced in \cite{de2014adjustment} and takes additional context into account: ticketing data when it is available, and knowledge of ridership on past courses. Comprehensive experiments are proposed to compare the different methods in Section~\ref{sec:experiments}.

\section{PROPOSED METHOD}
\label{sec:proposed_method}

\subsection{General principle}

The proposed method aims at denoising APC data to make it reliable using a constrained integer linear optimization algorithm. Our approach extends the one presented by de Oña \textit{et al.} \cite{de2014adjustment} to integrate additional data at the optimization stage: historical APC data to compute priors and ticketing data that defines a lower bound for denoised counts and is used to compute priors in combination with past counting data.

Let $y_i$ be the number of passengers boarding and $z_i$ the number of passengers alighting with $i$ the position of a station on the line. Let $O_i$ be the occupancy between station $i$ and station $i+1$ and $N$ be the number of stations on the line. For simplicity, $x_i$ denotes the ensemble of the $y_i$ and $z_i$, and $\mathcal{X}$ the set of $x_i$ on a course {i.e.} all the boarding and alighting counts of a course.
As boardings and alightings represent real public transport passengers, denoised counts must be integer values. Our objective is to find, from a set of observed counts $\mathcal{X}^\text{obs}$, a set of denoised counts $\mathcal{X}$ that respects a list of constraints imposed by public transport operational conditions and that optimizes the chosen objective function. When available, the set of ticketing counts $\mathcal{X}^V$ may be used as detailed in the following. The knowledge of the complete $\mathcal{X}^V$ set may require trip reconstruction as only boarding data is usually available \cite{hussain2021transit,trepanier2007individual,yan2019alighting,assemi2020improving,luca2021survey}. In this work, we use the statistical reconstruction proposed in~\cite{amblard2023bayesian} to get complete ticketing information for each course.

Constraints, objective function and optimization process are described in the following sections.

\subsection{Constraints} 

To ensure counts are consistent with operational constraints and passenger flow conservation, the denoised counts ${x}_{i}$ must respect the following constraints:

\begin{itemize}
\item total number of boardings and total number of alightings must be equal for each course:
\begin{equation}
\label{eq:boarding_equal_alighting}
    \sum_{i=1}^N{y}_{i} = \sum_{i=1}^N{z}_{i}
\end{equation}
\item boarding, alighting and occupancy must be positive and lower than a given capacity $L_{\text{max}}$ that can be the vehicle capacity or a factor of it:
\begin{equation}
\forall i, \quad
\begin{cases}
\label{eq:occupancy_positive}
    0 \leq {x}_{i} \leq L_{\text{max}} \\
    0 \leq {O}_{i} \leq L_{\text{max}}
\end{cases}
\end{equation}

\item there cannot be any boarding at the last stop, or any alighting at the first stop:
\begin{equation}
\begin{dcases}
\label{eq:no_boarding_at_last_stop_or_alighting_at_first_stop}
    y_{N} = 0 \\
    z_1 = 0
\end{dcases}
\end{equation}

\item computed boardings, alightings and occupancy must remain superior to ticketing data where available:
\begin{equation}
\label{eq:denoised_superior_to_ticketing}
    \forall i, \quad {x}_{i} \geq {x}_{i}^V .
\end{equation}
\end{itemize}

\subsection{Triangular similarity function} 

\begin{figure}
  \centering
  \includegraphics[width=0.7\linewidth]{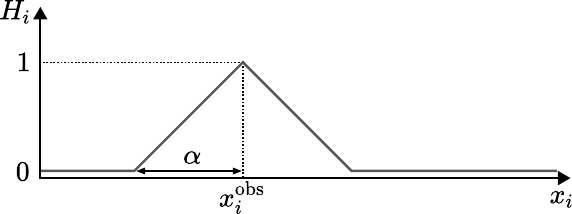}
  \caption{Triangular similarity function $H_i$. The function is characterized by the following two parameters: $x_i^\text{obs}$ the value of the observed count, and $\alpha$ the half margin.}
  \label{fig:triangle_cost_function}
\end{figure}

Fig.~\ref{fig:triangle_cost_function} shows the chosen similarity function, that is used during the first two stages of the optimization algorithm, to evaluate the proximity between denoised counts ${x}_{i}$ and observed counts ${x}_{i}^\text{obs}$. As indicated in \cite{de2014adjustment}, the choice of a triangular similarity function is relevant both from a computational perspective (piecewise linear function) and from the nature of the optimization problem addressed here. 
For each observation, the chosen similarity function writes 

\begin{equation}
\label{eq:triangular_similarity_function}
    H_i(x_i) = \max\left(0,1- \frac{\left|{x}_{i}-{x}_{i}^\text{obs}\right|}{\alpha}\right)
\end{equation}
with $\alpha$ the half margin {i.e.} the maximal allowed distance between denoised and observed counts before $H_i$ goes to zero.

Fig.~\ref{fig:step_1},~\ref{fig:step_2} and~\ref{fig:step_3} illustrate the three stages of the optimization algorithm: a count is represented there by a vertical bar on the similarity function associated with the observed count. Thus, the further the vertical bar is from the top of the triangle, the further the denoised count considered is from the observed count, and if the vertical bar is at the left (respectively at the right) of the top of the triangle, the denoised count considered is inferior (respectively superior) to the observed count.

\subsection{Three stage optimization algorithm}

\subsubsection{Stage I - Maximize the minimal similarity between denoised counts and observed counts}

For a course, among all the solutions that respect every constraint listed above, the first stage seeks to get rid of the outliers, {i.e.} solutions with a denoised count far away from the reference observed count. In this stage, the focus is for each solution on the count for which the similarity is minimal. The proposed algorithm aims at maximizing this minimal similarity. Therefore, the problem writes

\begin{equation}
\label{eq:optimization_step_1}
    \max_{x\in\mathcal X} \left(\min_{i}H_i\left(x_i\right)\right).
\end{equation}

\begin{figure}
  \centering
  \includegraphics[width=1\linewidth]{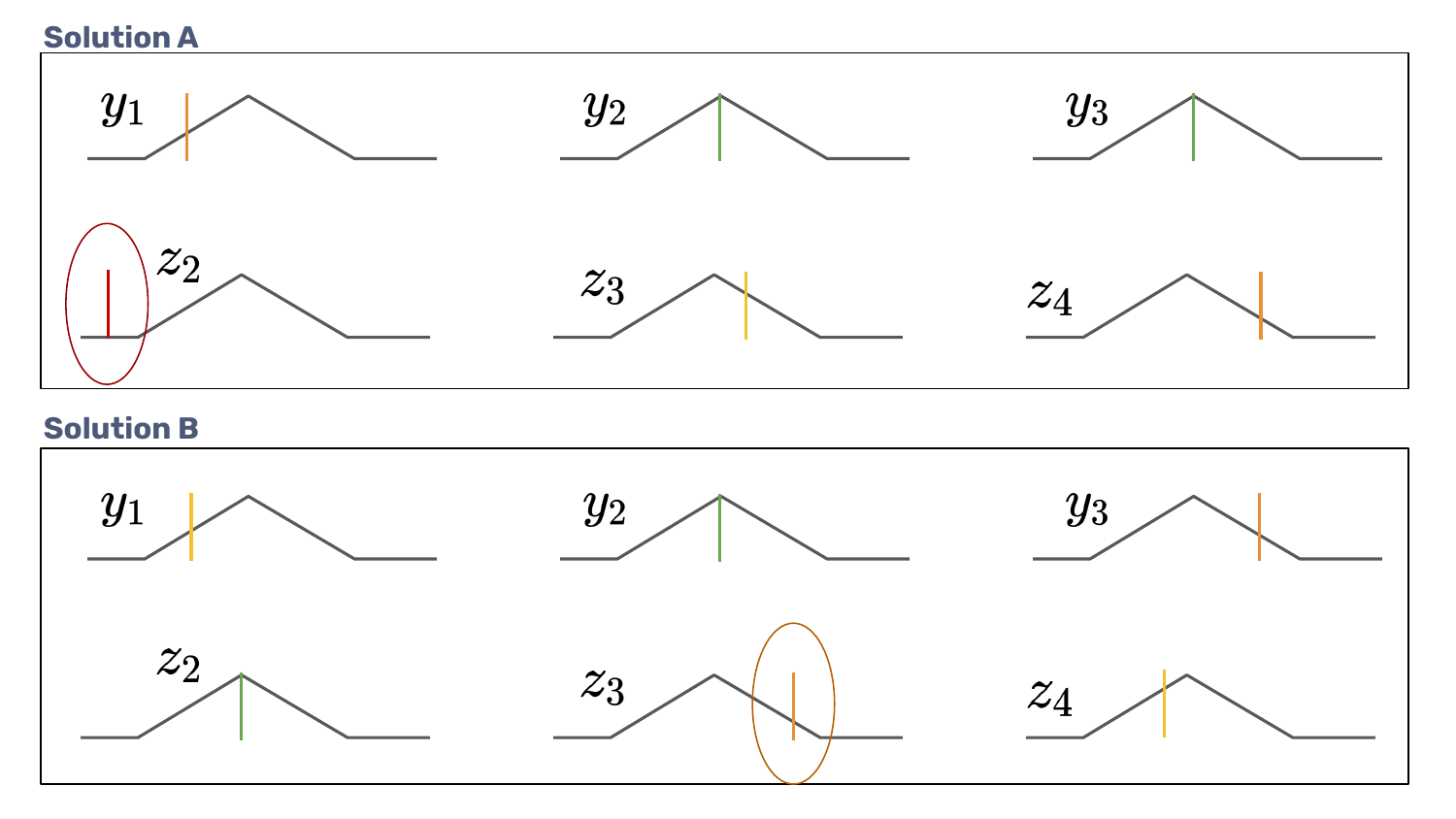}
  \caption{\textbf{Step I: Maximize the minimal similarity between denoised counts and observed counts.} On solution A, alighting 2 is the count with the maximal difference between the observed and denoised count. On solution B, it is the alighting 3. Solution A is less optimal than solution B because it has a smaller minimal similarity between observed and denoised counts.}
  \label{fig:step_1}
\end{figure}

Fig.~\ref{fig:step_1} illustrates how a solution is preferred to another based on this first stage. The second stage is built on top on this, namely the optimum of the first stage must be fulfilled before optimizing the objective function of the second stage.

\begin{figure}
  \centering
  \includegraphics[width=1\linewidth]{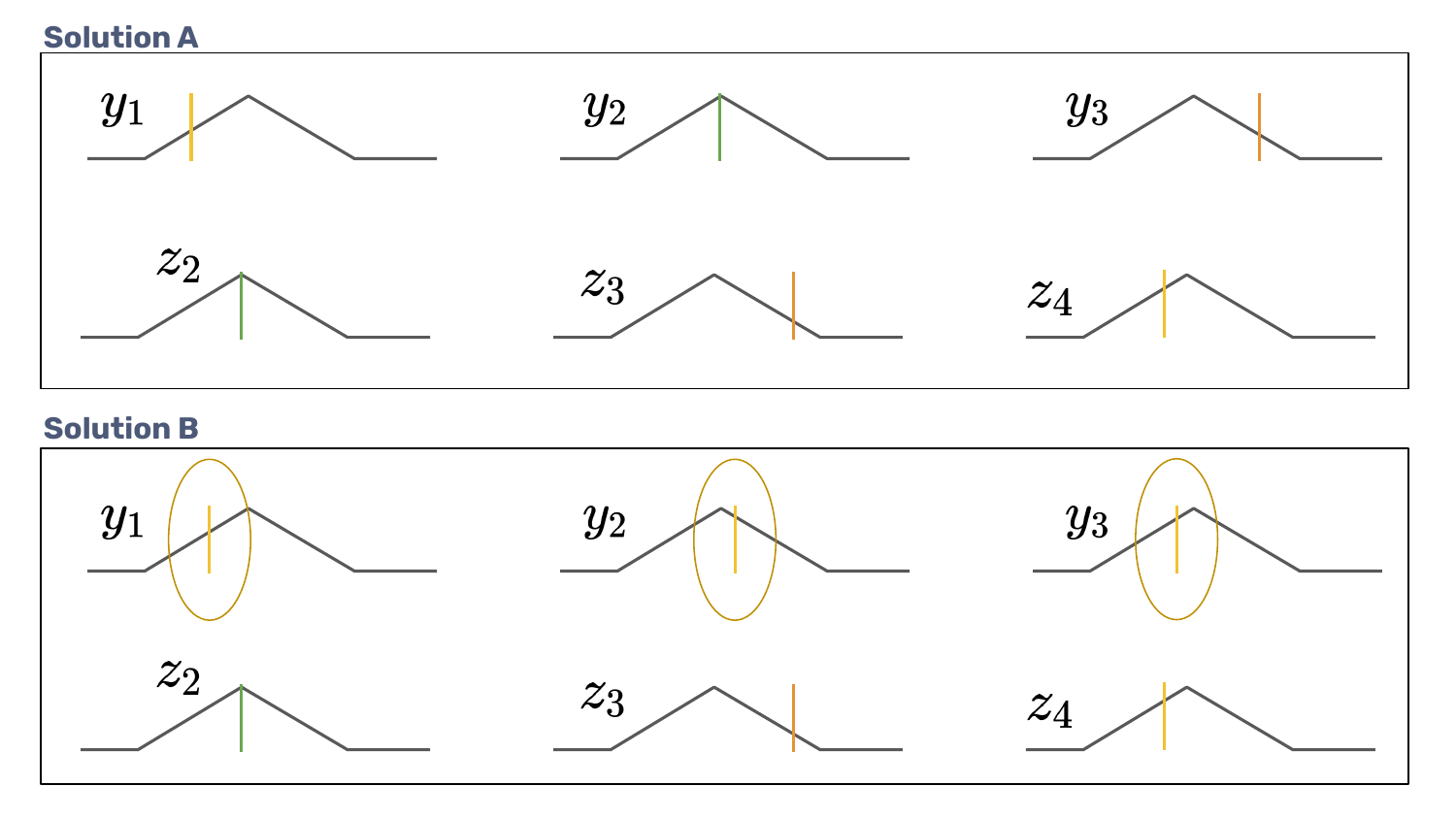}
  \caption{\textbf{Stage II: Maximize the sum of similarities between denoised and observed counts.} Between solutions A and B, only denoised boarding counts differ as denoised alighting counts are the same. Since the sum of similarities between denoised and observed counts is higher for solution B, solution A is less optimal than solution B.}
  \label{fig:step_2}
\end{figure}

\begin{figure}
  \centering
  \includegraphics[width=0.8\linewidth]{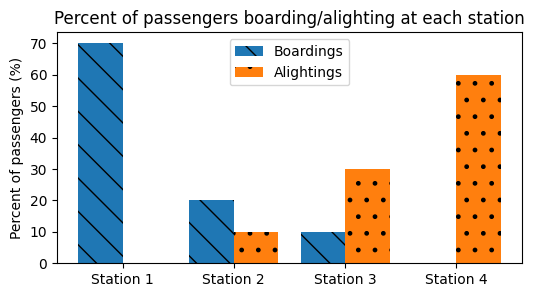}
  \caption{Example of priors for a given line and direction. 70\% of passengers board at station 1 in average. 30\% of passengers alight at station 3 in average.}
  \label{fig:priors}
\end{figure}

\begin{figure}
  \centering
  \includegraphics[width=1\linewidth]{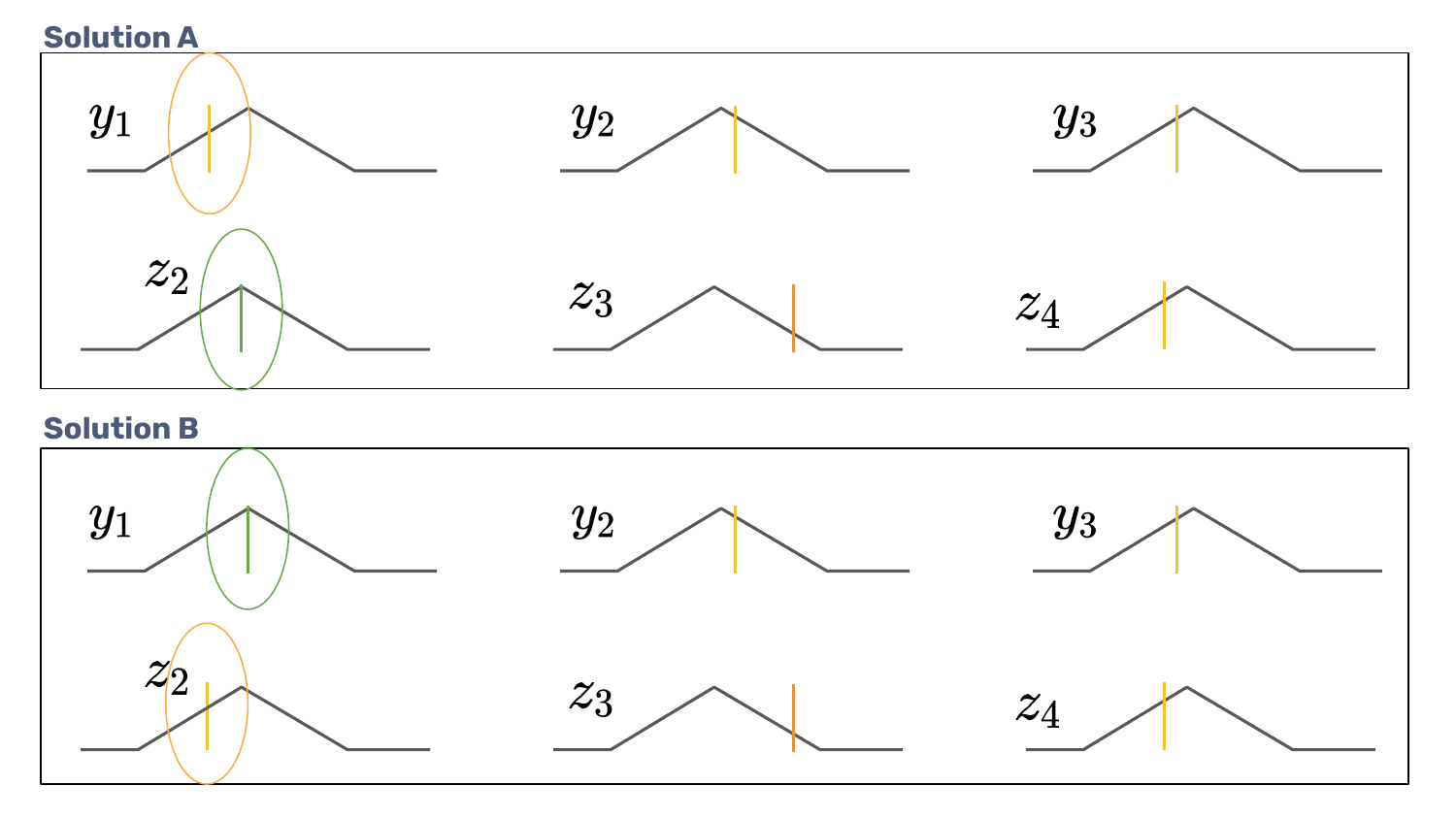}
  \caption{\textbf{Stage III: Choose the solution the closest to priors computed from historical data.} Solution B has more boardings at station 1 and more alightings at station 2 than solution A and is therefore closer to priors shown in Fig.~\ref{fig:priors}, as the rest of the counts is similar between both solutions. Solution B is considered to be the best solution among A and B.}
  \label{fig:step_3}
\end{figure}

\subsubsection{Stage II - Maximize the sum of similarities between denoised and observed counts}

The second stage consists in minimizing the distances between observed and denoised counts to find the solutions that are the closest to the observed counts. To this end, this stage focuses on the sum of similarities between denoised and observed counts. The algorithm considers as optimal the set of solutions for which this sum is maximal. Under the constraint of having reached the optimality of stage I, stage II writes

\begin{equation}
\label{eq:optimization_step_2}
    \max_\mathcal X \sum_{i}H_i(x_i).
\end{equation}
This quantity can be interpreted as an evaluation of observed counts quality: the higher the distance between the observed and the denoised counts, the lower the quality of the observed counts, since many changes are necessary to respect the constraints. This stage may recall an optimization with $\ell_1$ norm but the proposed approach uses a triangular similarity function instead of the $\ell_1$ distance.

Fig.~\ref{fig:step_2} illustrates how a solution is preferred to another based on this second step.

\subsubsection{Stage III - Use priors to choose among remaining solutions}

At this stage, several optimal solutions may coexist, with the same similarity to observed counts. The third step of the proposed approach selects, under the constraint of having reached the optimality of stage II, the closest solution to priors computed from historical data from counting cells and/or ticketing as detailed in Section~\ref{sec:priors}. For a given line and direction, priors take the form of percentages of boardings and alightings at each station over the total ridership of the line and direction. Fig.~\ref{fig:priors} illustrates an example of priors.

This stage minimizes the distance (in absolute value) between denoised counts and prior counts. It writes
\begin{equation}
\label{eq:optimization_step_3}
    \min_\mathcal X \sum_i \left| x_i - S \, p_{x_i}\right|
\end{equation}  
with $S=\sum_i y_i = \sum_i z_i$ the total (denoised) number of passengers, $x_i$ the boarding/alighting denoised counts, and $p_i$ the percentage of passengers that usually board/alight at station $i$ according to prior data.

Based on example priors shown in Fig.~\ref{fig:priors}, Fig.~\ref{fig:step_3} illustrates how the third step of the proposed optimization algorithm chooses between two solutions. 

\subsection{Priors computation}
\label{sec:priors}

Priors are computed using data from other courses on the same line and direction as the course to denoise, provided they also have APC and possibly ticketing data. No assumption is made here on the number of available courses in the history necessary to compute precise priors.

Priors take the form of proportions of boardings and alightings at each station, for a given line and direction. Boardings and alightings are treated separately but in similar ways. Proportions of passengers boarding (resp. alighting) at each station are computed over the entire set of courses with data on the same line and direction. If ticketing data is available, priors are additionally computed based on ticketing boardings and alightings. Finally, priors $p_{y_i}$ (resp. $p_{z_i}$) for boardings (resp. alightings) at a station $i$ write as a weighted average of APC priors $p_{y_i}^C$ (resp. $p_{z_i}^C$) and ticketing priors $p_{y_i}^V$ (resp. $p_{y_i}^V$):

\begin{equation}
\label{eq:priors_computation}
    p_{y_i} = r \,p_{y_i}^C + (1-r)\,p_{y_i}^V
\end{equation}
with $r$ the proportion of courses in the available history equipped with APC systems on the studied line and direction.

\section{EXPERIMENTS}
\label{sec:experiments}

\subsection{Experimental setup}

We set for the experiments the half margin of the similarity function $\alpha = \max\left(5, \frac{x_i^{\text{obs}}}{2}\right)$, chosen experimentally and from expert knowledge looking at specific real courses. We also authorize up to 140\% of the vehicle capacity as upper bound $L_{\text{max}}$ for the counts and occupancies (a vehicle can be overloaded).

After experimenting with a threshold on the number of courses on a route to compute priors, we decided not to enforce any threshold and to compute priors whatever the service frequency on the route. This permits to avoid random choices among optimal solutions after stage II of the optimization, even with a low coverage of the route with APC systems. We use a month of historical APC and ticketing data to compute priors.

Optimization may be infeasible for some courses: if ticketing data is greater than the capacity $L_{\text{max}}$ for instance. In this case, optimization is retried without the ticketing constraints, considering ticketing data is aberrant. 
Besides, the proposed optimization problem must be linearized to be solvable by linear solvers: to this end, an upper bound is required for the $H_i$ variables \cite{williams2013model}, therefore constraining as well the $x_i^{\text{obs}}$ counts. If aberrant counts are present in a course, the problem is infeasible in its linear form. The upper bound can either be set to a high value, which slows down the optimization, or to a reasonable value, which is the choice made here: we consider the counts cannot be higher than twice the capacity $L_{\text{max}}$.

\subsubsection{Baselines}

In the following, the proposed optimization method is evaluated against several other denoising alternatives:
\begin{itemize}
    \item a single optimization step, with $\ell_1$ norm between observed and denoised counts as cost function;
    \item a single optimization step, with $\ell_2$ norm between observed and denoised counts as cost function;
    \item the original optimization based denoising from \cite{de2014adjustment}, with the two same optimization steps but without the ticketing constraints and the priors;
    \item the method proposed in \cite{amblard2023bayesian} using a Gibbs sampling: the principle is to start from a valid sequence of counts and iteratively adjust each count by sampling from the posterior distribution derived from the observed counts.
\end{itemize}

These approaches have been chosen for their simplicity ($\ell_1$ and $\ell_2$), the proximity to our method (\cite{de2014adjustment}), and an approach completely different from integer optimization (\cite{amblard2023bayesian}). 
The integer linear optimization problems are solved using the open source COIN-OR CBC solver (release 2.10.10), and the quadratic one using the free version of CPLEX available for small optimization problems (release 22.1.1). Gibbs sampling is implemented in Python.

\subsubsection{Real and simulated datasets}

To evaluate the accuracy of the proposed method, we compare the result of denoising to manual counts on the field conducted on a public transportation network named Network A in the following (the name of the network is anonymized at the operator's demand). 74 records are available over three lines. For each record, the number of boarding and alighting passengers at the station is recorded as well as the occupancy onboard the vehicle. This permits on the one hand to assess the quality of counting cells data with respect to the manual counts, and on the other hand to evaluate the improvement provided by the proposed denoising method.

To complete performance evaluation, we simulate extensive scenarios where counting cells data can be under or over-estimated, contain outliers, or just present a Gaussian noise. To this end, we take boarding and alighting ticketing data from the Angers network and select eight courses from two different lines (tramway and bus) with varying occupancy rates. Ticketing counts are then distorted according to the desired bias. Counts can be overestimated (randomly adding between 1 and 5 passengers), underestimated (randomly removing between 1 and 5 passengers), or overestimated at low counts and underestimated at high counts (using a linear slope of coefficient 0.3). Outliers can be randomly assigned (putting a random count either to twice the capacity or to 0). In addition, a Gaussian noise of 10\% is systematically applied. 

Finally, we conducted experiments on data from four public transportation networks of cities in France: Angers, Nevers, Brest that have both APC and ticketing data, in addition to the anonymized Network A with its manual counts. No baseline is available for these first three networks but course examples are studied and the impact on measured counts evaluated.

\subsection{Computation time}

Table~\ref{tab:computation_time} presents the computation times for each alternative method in milliseconds, averaged over 310 courses. Computations were performed on Intel x86 architecture single-thread, although they are easily parallelizable over the courses.
The computation time of the proposed method using either CBC or CPLEX as solver is also evaluated: CPLEX is significantly faster but we retain CBC as an open source solver. $\ell_2$ minimization takes a significantly longer amount of time because of its non-linearity, although solved by CPLEX. Adding several stages of optimization slows down the computation, but the proposed method still runs faster than the statistical one from \cite{amblard2023bayesian}, while not adding a significant overload to the baseline from \cite{de2014adjustment}.

\begin{table}
    \centering
    \caption{Computation time in milliseconds for one course for each proposed method}
    \label{tab:computation_time}
    \begin{tabular}{lc}
    \toprule
         Denoising method & Computation time per course (ms) \\
         \midrule
         $\ell_1$ (CBC) & 39 \\
         $\ell_2$ (CPLEX) & 33600 (*) \\
         \cite{de2014adjustment} (CBC) & 225 \\
         \cite{amblard2023bayesian} & 1200 (†) \\
         \midrule
         Ours (CBC) & 296 \\
         Ours (CPLEX) & 81 \\
    \bottomrule
    \end{tabular}
    \\ (*) With a timeout of 60 seconds per course
    \\ (†) Computation time is quadratic with the highest possible count,\\ therefore highly depends on the network's use
\end{table}

\subsection{Global evaluation with manual counts and simulations}

\begin{table}
    \centering
    \caption{Mean absolute difference between manual counts and counts measured by counting cells}
    \label{tab:cc_quality}
    \begin{tabular}{lccc}
    \toprule
         Dataset & Boardings & Alightings & Occupancy \\
         \midrule
         Network A & 11.11 & 8.74 & 60.15 \\
         \midrule
         Gaussian noise & 0.36 & 0.30 & 4.88 \\
         Overestimated & 4.23 & 4.06 & 34.45 \\
         Underestimated & 1.85 & 1.52 & 13.92 \\
         Over/under & 9.92 & 10.15 & 19.41 \\
         Outliers & 3.33 & 3.17 & 66.79 \\
    \bottomrule
    \end{tabular}
\end{table}

Table~\ref{tab:cc_quality} depicts the mean absolute differences between observed counts and true counts (measured on the field and compared to real APC data for Network A, true counts compared to simulated counts for simulations). The error on occupancy is significantly higher than the errors on boardings and alightings for all scenarios. The plausible interpretation of these results is that counting cells data have proper quality, however the combination of observed boarding and alighting counts along with their associated imprecisions leads to absurd occupancies. It is worth noting that this effect is particularly important for the simulation that includes outliers and for the real data from Network A. Therefore, denoising is expected to significantly lower errors on occupancies, rather than those on boardings and alightings.

\begin{table}
    \centering
    \caption{Occupancy mean absolute error of counting cells (baseline) and denoising methods, for different datasets}
    \label{tab:occupancy_mae}
    \begin{tabular}{lccccc}
    \toprule
         Dataset & Baseline & $\ell_1$ & $\ell_2$ & \cite{amblard2023bayesian} & Ours \\
         \midrule
         Network A & 60.15 & 50.05 & \textbf{46.95} & 48.09 & 47.18 \\
         \midrule
         Gaussian noise & 4.88 & 2.77 & 2.08 & 4.16 & \textbf{1.68} \\
         Overestimated & 34.45 & \textbf{8.86} & 9.83 & 15.45 & 10.30 \\
         Underestimated & 13.92 & 9.15 & \textbf{6.14} & 9.58 & 7.54 \\
         Over/under & 19.41 & 15.11 & \textbf{14.62} & 16.03 & 14.90 \\
         Outliers & 66.79 & 22.80 & 25.05 & 16.48 & \textbf{9.93} \\
         \midrule
         Rank sum & 30 & 17 & \textbf{11} & 21 & \textbf{11} \\
    \bottomrule
    \end{tabular}
\end{table}

Table~\ref{tab:occupancy_mae} compares the different denoising methods applied to the test datasets. The comparison with \cite{de2014adjustment} is not done here since neither ticketing data nor historical counting cells data is available to compute priors on. The objective here is to objectively compare several denoising techniques on datasets that have ground truth available.

First, errors on occupancies are indeed well corrected with any denoising method. The statistical model gets the worst performance in the majority of cases. Moreover, summing the ranks over the six datasets gives $\ell_2$ and the proposed method as equally best methods. $\ell_2$ minimization performs indeed well over various scenarios but cannot be retained for production purposes because of its high computation time. Finally, the proposed method is second best in most cases including the real Network A dataset. It is also the best to handle outliers compared to other approaches. In particular, $\ell_1$ and $\ell_2$ minimization methods get the highest errors, probably since they do not have a procedure to absorb outliers. This last claim is also examplified in the following section. 

\subsection{Real courses examples}
\label{sec:courses_examples}

\begin{figure}
    \centering
    \includegraphics[width=\linewidth]{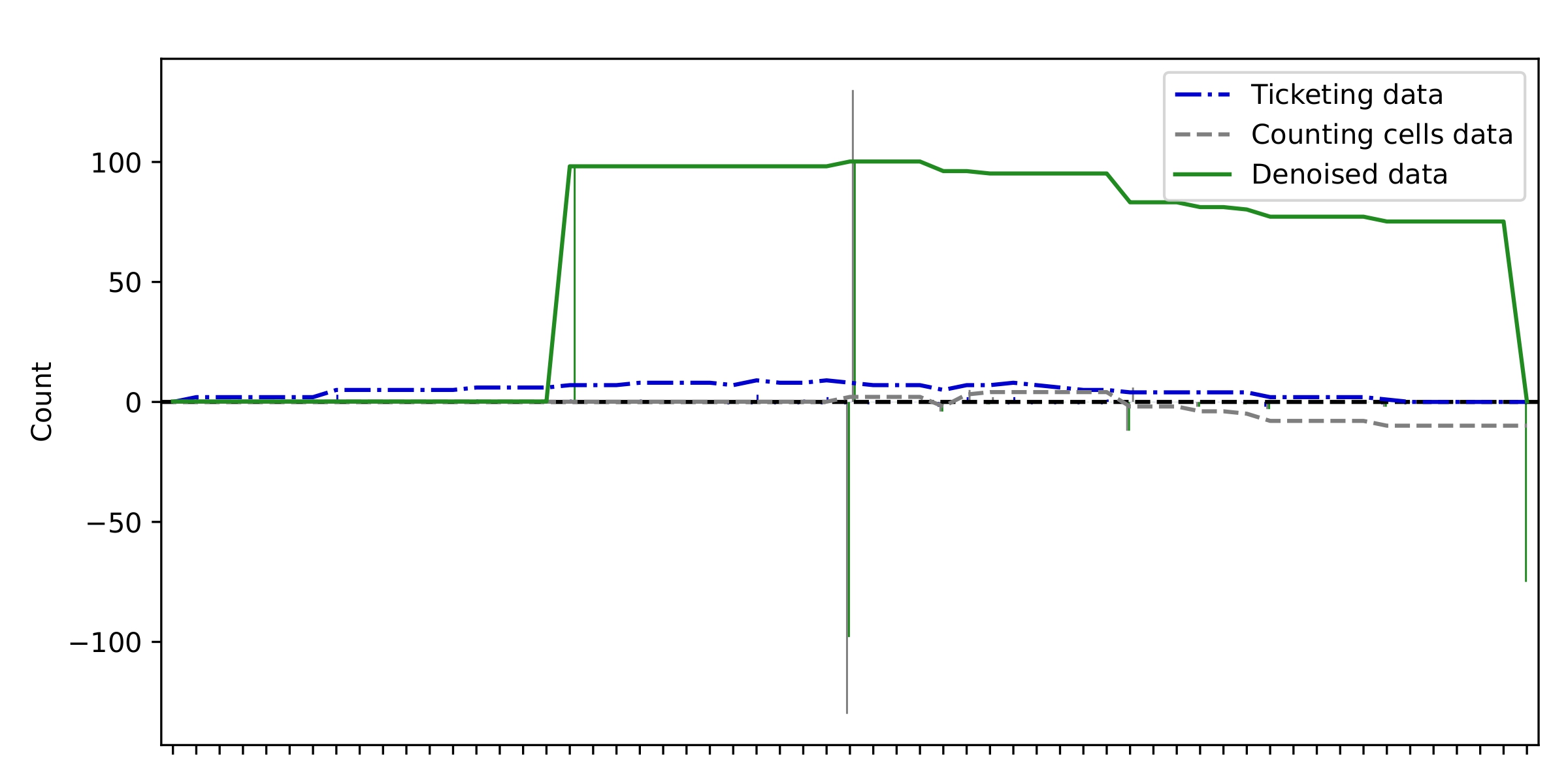}
    \includegraphics[width=\linewidth]{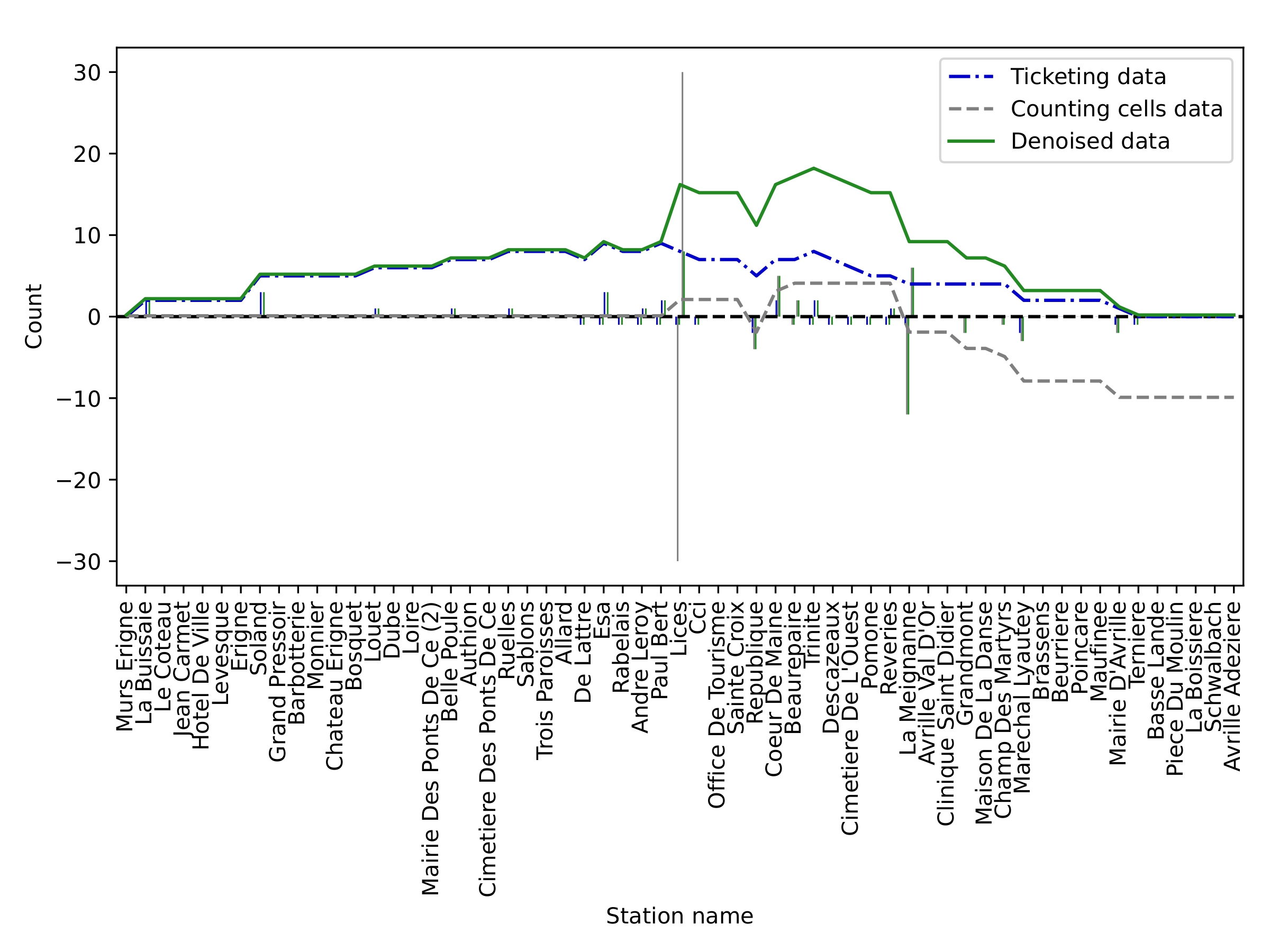}
    \caption{A course at Angers network. Top: denoised by $\ell_1$ method. Bottom: denoised by the proposed approach. Counting cells counts have been clipped to fit in the figure.}
    \label{fig:outliers_example}
\end{figure}

\begin{figure*}
    \centering
    \includegraphics[width=\textwidth]{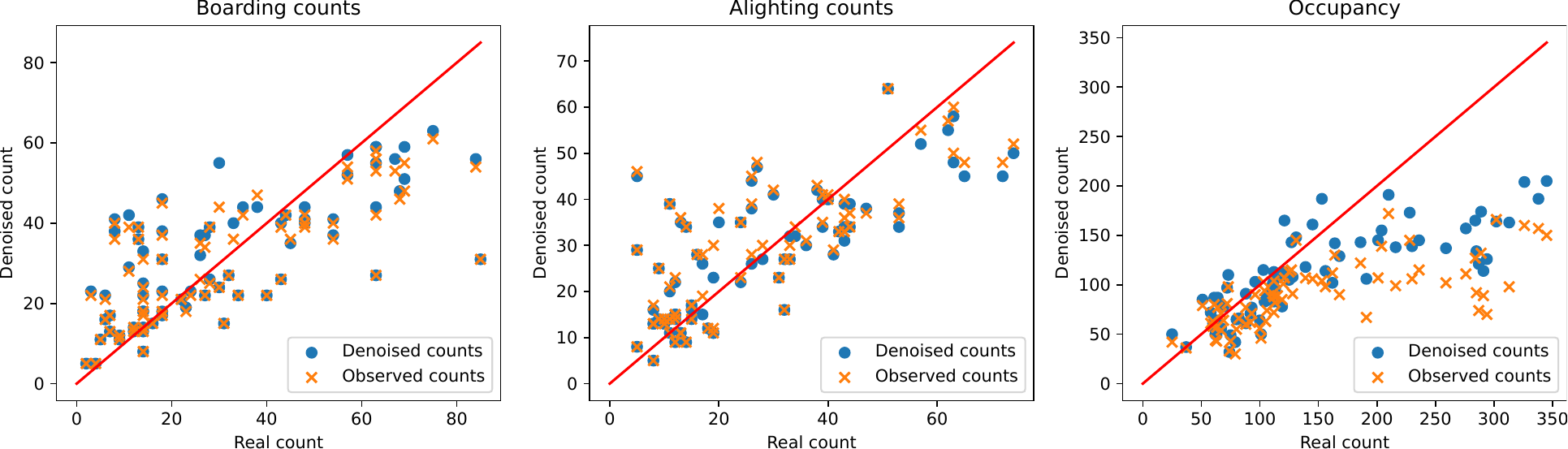}
    \caption{Boarding, alighting and occupancy measures in Network A: counting cells and denoised counts (horizontal axis) against manual counts (vertical axis).}
    \label{fig:networkA}
\end{figure*}

\begin{figure}
    \centering
    \includegraphics[width=\linewidth]{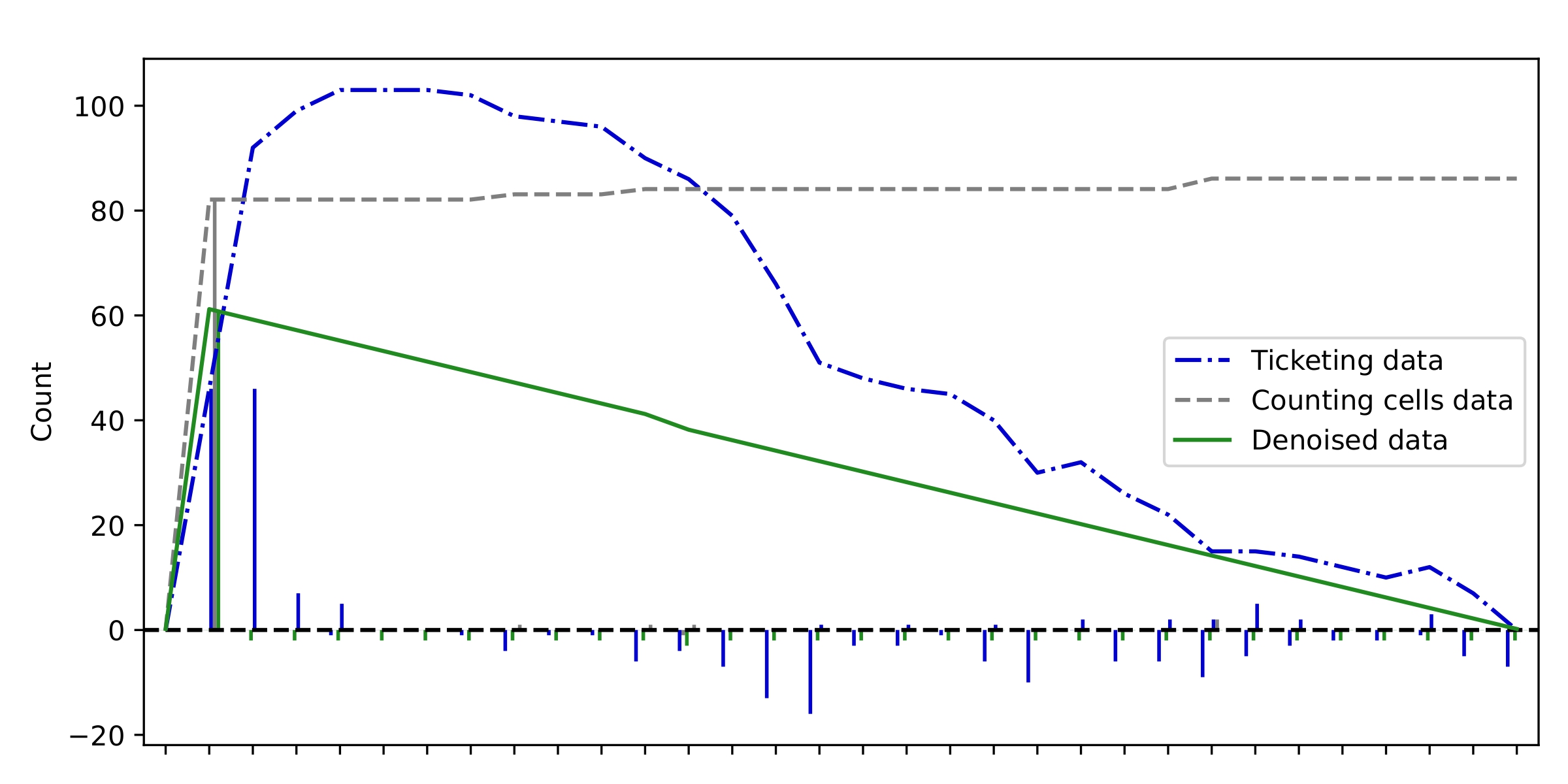}
    \includegraphics[width=\linewidth]{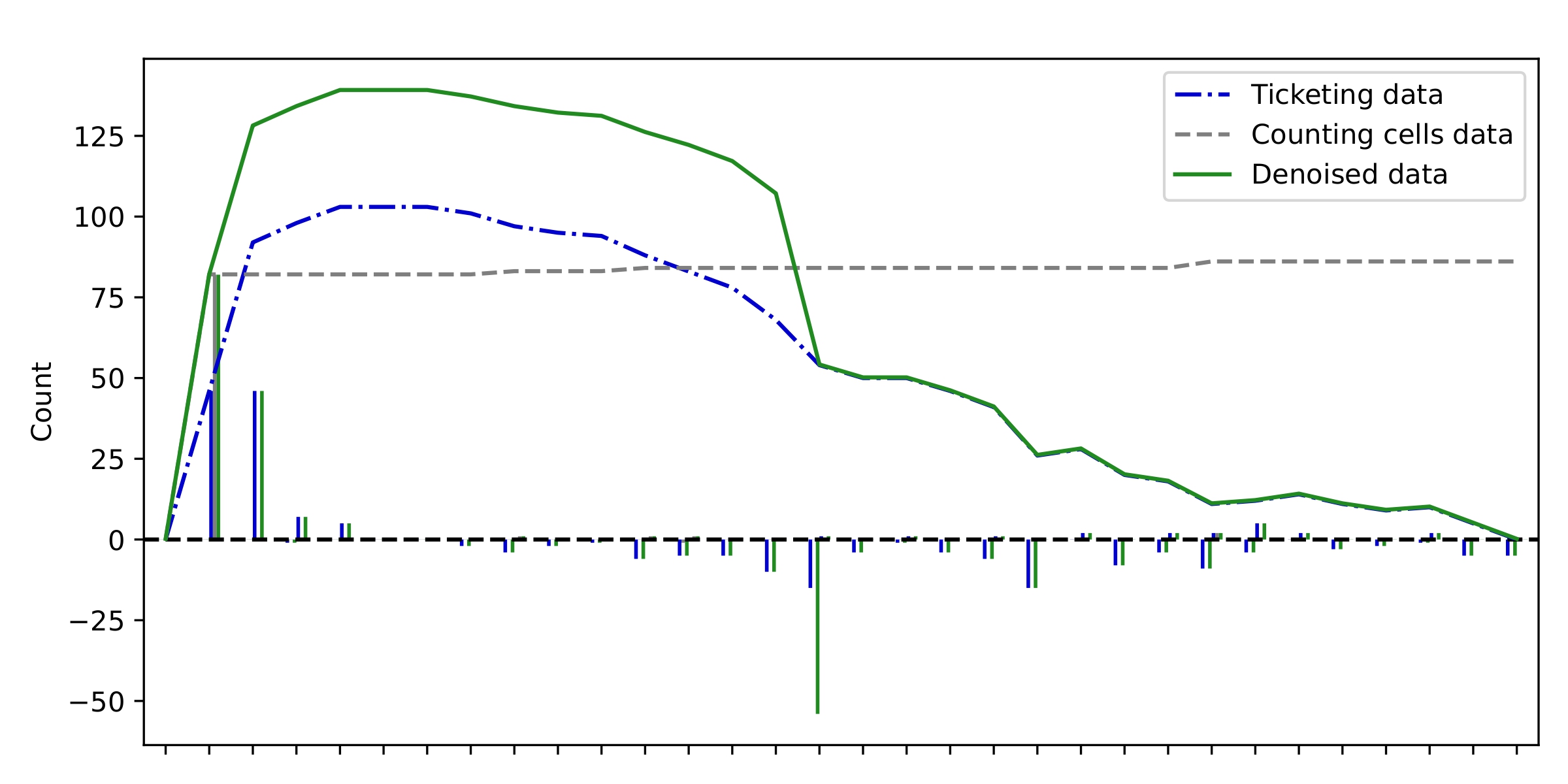}
    \includegraphics[width=\linewidth]{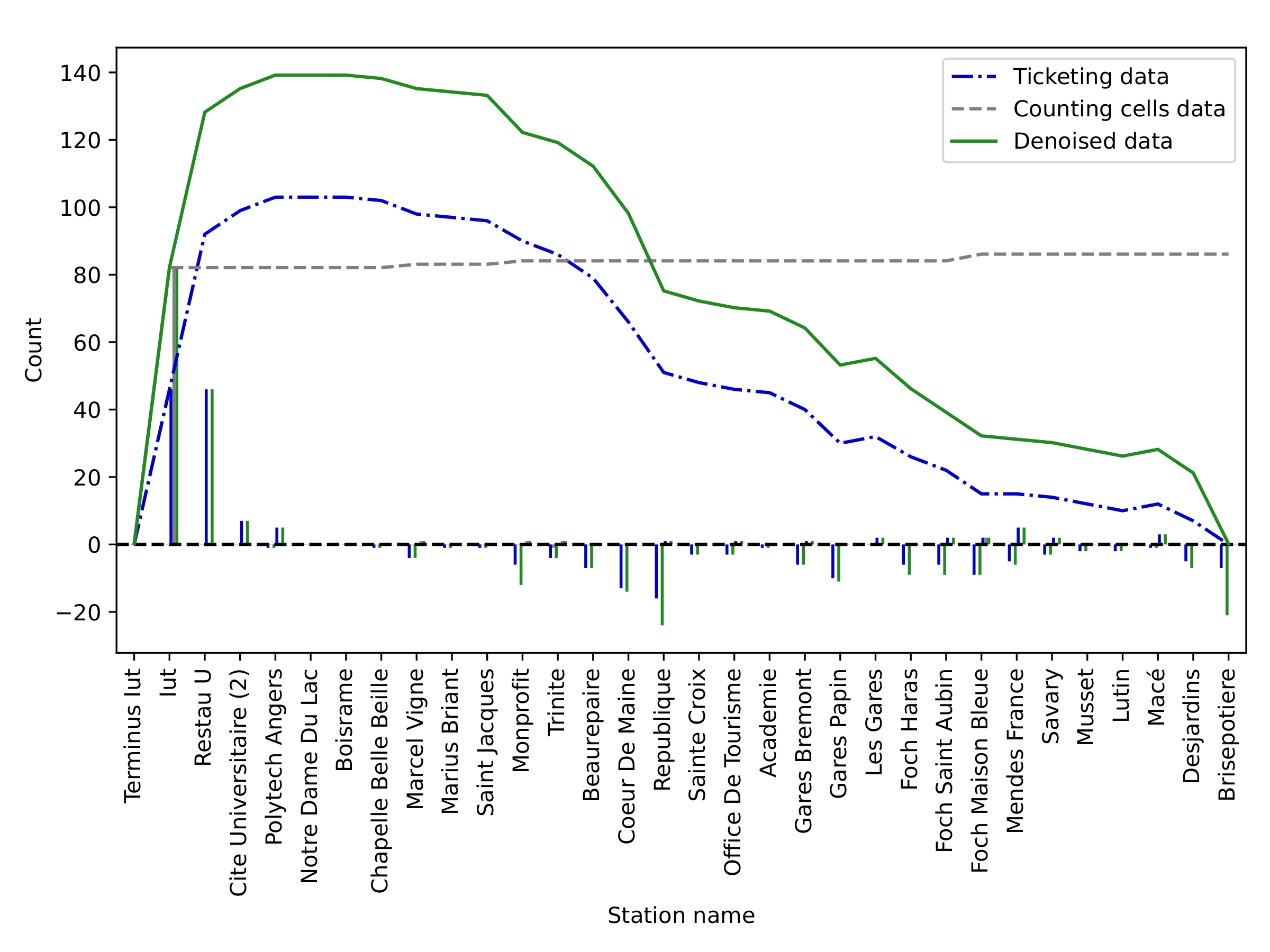}
    \caption{A course at Angers network. Top: denoised by method from \cite{de2014adjustment}. Middle: denoised by the proposed approach, without priors. Equivalent to adding ticketing constraints to \cite{de2014adjustment}. Bottom: denoised by the proposed approach, with priors.}
    \label{fig:ticketing_example}
\end{figure}

To further demonstrate the relevance of the proposed approach, a few courses are illustrated and discussed here. Fig.~\ref{fig:outliers_example} shows the relevance of the proposed approach compared to the simpler $\ell_1$ minimization method. The illustrated course presents outlier counts: at station ``Lices'', counting cells recorded 254 boarding and 252 alighting passengers, which is impossible since no one boarded on this course prior to this station. Even without ground truth on this subject, we can reasonably assume these counts are aberrant and should be ignored during denoising. The top plot on Fig.~\ref{fig:outliers_example} shows how denoising is performed by the $\ell_1$ method. Since this method minimizes the distances between observed and denoised counts, the result remains as close as possible to the observed (absurd) counts. In addition, boarding passengers have been added to a station earlier in the course in order to have the necessary occupancy to make passengers alight later at ``Lices'', and alighting passengers added to the last station of the course to empty the vehicle. In contrast, the proposed approach (bottom plot on Fig.~\ref{fig:outliers_example}) ignores the aberrant counts thanks to the triangular similarity function and outputs visibly more consistent denoised counts and occupancies.

Similarly, Fig.~\ref{fig:ticketing_example} compares denoising methods for a same course. Here, the relevance of ticketing constraints and of the third step using priors is illustrated. The chosen course has a high number of boarding passengers at ``Iut'', which does not seem aberrant considering the ticketing data. However, there is only one alighting passenger according to APC data, violating passenger flow conservation. On the top plot, the method from \cite{de2014adjustment} diminishes slightly this high boarding count and then dispatches the alighting passengers in a uniform manner over the remaining stations. Indeed, the baseline algorithm from \cite{de2014adjustment} has no information on where passengers are the most likely to alight. On the middle plot, denoising is aware of ticketing data and outputs boarding and alighting counts greater than ticketing counts. Missing alighting counts are all assigned at the same station ``République'', as if the observed alighting count at this station is aberrant. Finally, the proposed approach (bottom plot) assigns alighting counts more smoothly on the remaining stations, choosing carefully on which stations should the most people board or alight.


\subsection{Impact on counts}

Table~\ref{tab:impact_on_counts} illustrates the impact of denoising over measured counts for several public transportation networks of medium-size cities in France, using October 2022 data. While the changes to boarding and alighting counts themselves are minor, occupancy as a result is significantly affected. Moreover, occupancy is increased by denoising for all networks, corresponding to field knowledge that occupancy is underestimated by counting cells.

\begin{table}
    \centering
    \caption{Mean bias and absolute differences between measured and denoised counts}
    \label{tab:impact_on_counts}
    \begin{tabular}{llcc}
    \toprule
         & & Mean bias & Mean absolute delta \\
         \midrule
         Angers & Boardings & 0.38 & 0.56 \\
         & Alightings & 0.59 & 0.67 \\
         & Occupancy & 4.10 & 11.07 \\ 
         \midrule
         Nevers & Boardings & 0.21 & 0.25 \\
         & Alightings & 0.19 & 0.25 \\
         & Occupancy & 1.86 & 4.30 \\
         \midrule
         Brest & Boardings & 0.44 & 0.53 \\
         & Alightings & 0.51 & 0.61 \\
         & Occupancy & 3.52 & 6.60 \\
         \midrule
         Network A & Boardings & -0.19 & 0.68 \\
         & Alightings & -0.21 & 0.72 \\
         & Occupancy & 3.26 & 11.01 \\
    \bottomrule
    \end{tabular}
\end{table}

Fig.~\ref{fig:networkA} shows measured, manual and denoised counts for Network A, where manual counts are available. Counts seem to be overestimated at low values and overestimated at high values, similarly to the over/under dataset we simulated. While few differences are visible on boardings and alightings, occupancy is well rectified although not perfectly. It was indeed largely underestimated, especially at high values.

\section{CONCLUSIONS}
\label{sec:conclusion}

To make data from APC systems reliable and usable, we proposed a method of constrained optimization fixing the noise and biases in counting cells data. Our method relies on constraints from flow conservation and ticketing data that are enforced through three stages of optimization: stage I removes outliers, stage II maximizes the similarity between solutions and observed data and stage III chooses the closest solution to priors computed from both past ticketing and APC data. 

Upon evaluation on data from several networks in France, we showed that the proposed method reconciles satisfying performances with reasonable computation time, and is particularly relevant to handle outliers. In particular, the help of ticketing data and historical ticketing and APC data permits to obtain more robust and reliable results when studying individual courses. 

As future work, specific prior knowledge about APC systems biases could be taken into account: for instance, a session of manual counts could help identify and quantify recurring biases such as a systematic overestimation of alighting counts. This knowledge could be used to calibrate the triangular similarity function, whose center or slopes could be shifted. In addition, this denoising method can be used in combination with a subsequent algorithm for other applications, for instance occupancy forecasting or O/D matrices reconstruction, since denoised APC counts are more robust than the raw ones.

\section*{ACKNOWLEDGMENT}

We would like to thank the Irigo network (city of Angers), the Taneo network (city of Nevers) and the Bibus network (Brest) for allowing us to publish the results of our algorithms on their data. We also warmly thank the Citio team for providing us with working and computing infrastructure, clean and formatted data, and domain expert knowledge to help us analyze results.

\bibliographystyle{IEEEtran}
\bibliography{bibliography}

\begin{thebibliography}{10}
\providecommand{\url}[1]{#1}
\csname url@rmstyle\endcsname
\providecommand{\newblock}{\relax}
\providecommand{\bibinfo}[2]{#2}
\providecommand\BIBentrySTDinterwordspacing{\spaceskip=0pt\relax}
\providecommand\BIBentryALTinterwordstretchfactor{4}
\providecommand\BIBentryALTinterwordspacing{\spaceskip=\fontdimen2\font plus
\BIBentryALTinterwordstretchfactor\fontdimen3\font minus \fontdimen4\font\relax}
\providecommand\BIBforeignlanguage[2]{{%
\expandafter\ifx\csname l@#1\endcsname\relax
\typeout{** WARNING: IEEEtran.bst: No hyphenation pattern has been}%
\typeout{** loaded for the language `#1'. Using the pattern for}%
\typeout{** the default language instead.}%
\else
\language=\csname l@#1\endcsname
\fi
#2}}

\bibitem{kuchar2023passenger}
P.~Kuch{\'a}r, R.~Pirn{\'\i}k, A.~Janota, B.~Malobick{\`y}, J.~Kub{\'\i}k, and D.~{\v{S}}i{\v{s}}mi{\v{s}}ov{\'a}, ``Passenger occupancy estimation in vehicles: A review of current methods and research challenges,'' \emph{Sustainability}, vol.~15, no.~2, p. 1332, 2023.

\bibitem{grgurevic2022review}
I.~Grgurevi{\'c}, K.~Jur{\v{s}}i{\'c}, and V.~Raji{\v{c}}, ``Review of automatic passenger counting systems in public urban transport,'' in \emph{5th EAI International Conference on Management of Manufacturing Systems}.\hskip 1em plus 0.5em minus 0.4em\relax Springer, 2022, pp. 1--15.

\bibitem{comparisonWLAN}
T.~Madsen, H.-P. Schwefel, L.~Mikkelsen, and A.~Burggraf, ``Comparison of {WLAN} probe and light sensor-based estimators of bus occupancy using live deployment data,'' \emph{Sensors}, vol.~22, no.~11, p. 4111, 2022.

\bibitem{nasir2018automatic}
A.~Nasir, N.~Gharib, and H.~Jaafar, ``Automatic passenger counting system using image processing based on skin colour detection approach,'' in \emph{2018 international conference on computational approach in smart systems design and applications (ICASSDA)}.\hskip 1em plus 0.5em minus 0.4em\relax IEEE, 2018, pp. 1--8.

\bibitem{hsu2020estimation}
Y.-W. Hsu, Y.-W. Chen, and J.-W. Perng, ``Estimation of the number of passengers in a bus using deep learning,'' \emph{Sensors}, vol.~20, no.~8, p. 2178, 2020.

\bibitem{zhang2020tiny}
S.~Zhang, Y.~Wu, C.~Men, and X.~Li, ``Tiny {YOLO} optimization oriented bus passenger object detection,'' \emph{Chinese Journal of Electronics}, vol.~29, no.~1, pp. 132--138, 2020.

\bibitem{klauser2015tof}
D.~Klauser, G.~B{\"a}rwolff, and H.~Schwandt, ``A {TOF}-based automatic passenger counting approach in public transportation systems,'' in \emph{AIP Conference Proceedings}, vol. 1648, no.~1.\hskip 1em plus 0.5em minus 0.4em\relax AIP Publishing LLC, 2015, p. 850113.

\bibitem{amin2008automated}
I.~Amin, A.~J. Taylor, F.~Junejo, A.~Al-Habaibeh, and R.~M. Parkin, ``Automated people-counting by using low-resolution infrared and visual cameras,'' \emph{Measurement}, vol.~41, no.~6, pp. 589--599, 2008.

\bibitem{choi2017bi}
J.~W. Choi, X.~Quan, and S.~H. Cho, ``Bi-directional passing people counting system based on {IR-UWB} radar sensors,'' \emph{IEEE Internet of Things Journal}, vol.~5, no.~2, pp. 512--522, 2017.

\bibitem{nitti2020iabacus}
M.~Nitti, F.~Pinna, L.~Pintor, V.~Pilloni, and B.~Barabino, ``iabacus: A wi-fi-based automatic bus passenger counting system,'' \emph{Energies}, vol.~13, no.~6, p. 1446, 2020.

\bibitem{patlins2015new}
A.~Patlins and N.~Kunicina, ``The new approach for passenger counting in public transport system,'' in \emph{2015 IEEE 8th International Conference on Intelligent Data Acquisition and Advanced Computing Systems: Technology and Applications (IDAACS)}, vol.~1.\hskip 1em plus 0.5em minus 0.4em\relax IEEE, 2015, pp. 53--57.

\bibitem{ben1985alternative}
M.~Ben-Akiva, P.~P. Macke, and P.~S. Hsu, \emph{Alternative methods to estimate route-level trip tables and expand on-board surveys}, 1985, no. 1037.

\bibitem{lamond1981bregman}
B.~Lamond and N.~F. Stewart, ``Bregman's balancing method,'' \emph{Transportation Research Part B: Methodological}, vol.~15, no.~4, pp. 239--248, 1981.

\bibitem{ji2015transit}
Y.~Ji, R.~G. Mishalani, and M.~R. McCord, ``Transit passenger origin--destination flow estimation: Efficiently combining onboard survey and large automatic passenger count datasets,'' \emph{Transportation Research Part C: Emerging Technologies}, vol.~58, pp. 178--192, 2015.

\bibitem{furth2005making}
P.~G. Furth, J.~G. Strathman, and B.~Hemily, ``Making automatic passenger counts mainstream: Accuracy, balancing algorithms, and data structures,'' \emph{Transportation research record}, vol. 1927, no.~1, pp. 206--216, 2005.

\bibitem{barabino2014offline}
B.~Barabino, M.~Di~Francesco, and S.~Mozzoni, ``An offline framework for handling automatic passenger counting raw data,'' \emph{IEEE Transactions on Intelligent Transportation Systems}, vol.~15, no.~6, pp. 2443--2456, 2014.

\bibitem{van1982consistent}
H.~J. Van~Zuylen and D.~M. Branston, ``Consistent link flow estimation from counts,'' \emph{Transportation Research Part B: Methodological}, vol.~16, no.~6, pp. 473--476, 1982.

\bibitem{amblard2023bayesian}
V.~Amblard, A.~Dib, N.~Cherrier, and G.~Barthe, ``A {Bayesian} {Markov} model for station-level origin-destination matrix reconstruction,'' in \emph{Machine Learning and Knowledge Discovery in Databases: European Conference, ECML PKDD 2022, Grenoble, France, September 19--23, 2022, Proceedings, Part VI}.\hskip 1em plus 0.5em minus 0.4em\relax Springer, 2023, pp. 538--553.

\bibitem{yin2017l1}
P.~Yin, Z.~Sun, W.-L. Jin, and J.~Xin, ``l1-minimization method for link flow correction,'' \emph{Transportation Research Part B: Methodological}, vol. 104, pp. 398--408, 2017.

\bibitem{vanajakshi2004loop}
L.~Vanajakshi and L.~Rilett, ``Loop detector data diagnostics based on conservation-of-vehicles principle,'' \emph{Transportation research record}, vol. 1870, no.~1, pp. 162--169, 2004.

\bibitem{kikuchi2006method}
S.~Kikuchi, S.~Mangalpally, and A.~Gupta, ``Method for balancing observed boarding and alighting counts on a transit line,'' \emph{Transportation research record}, vol. 1971, no.~1, pp. 42--50, 2006.

\bibitem{de2014adjustment}
J.~de~O{\~n}a, P.~G{\'o}mez, and E.~M{\'e}rida-Casermeiro, ``Adjustment boarding and alighting passengers on a bus transit line using qualitative information,'' \emph{Applied Mathematical Modelling}, vol.~38, no.~3, pp. 1147--1158, 2014.

\bibitem{hussain2021transit}
E.~Hussain, A.~Bhaskar, and E.~Chung, ``Transit {OD} matrix estimation using smartcard data: Recent developments and future research challenges,'' \emph{Transportation Research Part C: Emerging Technologies}, vol. 125, p. 103044, 2021.

\bibitem{trepanier2007individual}
M.~Tr{\'e}panier, N.~Tranchant, and R.~Chapleau, ``Individual trip destination estimation in a transit smart card automated fare collection system,'' \emph{Journal of Intelligent Transportation Systems}, vol.~11, no.~1, pp. 1--14, 2007.

\bibitem{yan2019alighting}
F.~Yan, C.~Yang, and S.~V. Ukkusuri, ``Alighting stop determination using two-step algorithms in bus transit systems,'' \emph{Transportmetrica A Transport Science}, vol.~15, no.~2, pp. 1522--1542, 2019.

\bibitem{assemi2020improving}
B.~Assemi, A.~Alsger, M.~Moghaddam, M.~Hickman, and M.~Mesbah, ``Improving alighting stop inference accuracy in the trip chaining method using neural networks,'' \emph{Public Transport}, vol.~12, no.~1, pp. 89--121, 2020.

\bibitem{luca2021survey}
M.~Luca, G.~Barlacchi, B.~Lepri, and L.~Pappalardo, ``A survey on deep learning for human mobility,'' \emph{ACM Computing Surveys (CSUR)}, vol.~55, no.~1, pp. 1--44, 2021.

\bibitem{williams2013model}
H.~P. Williams, \emph{Model building in mathematical programming}.\hskip 1em plus 0.5em minus 0.4em\relax John Wiley \& Sons, 2013.

\end{thebibliography}

\end{document}